\documentclass[sigconf]{acmart}
\AtBeginDocument{%
  }

\copyrightyear{2026}
\acmYear{2026}
\setcopyright{cc}
\setcctype{by}
\acmConference[SA Technical Communications '26]{SIGGRAPH Asia 2026 Technical Communications}{December 01--04, 2026}{Kuala Lumpur, Malaysia}
\acmBooktitle{SIGGRAPH Asia 2026 Technical Communications (SA Technical Communications '26), December 01--04, 2026, Kuala Lumpur, Malaysia}
\acmDOI{10.1145/3829339.3847840}
\acmISBN{979-8-4007-2841-9/2026/12}

\acmSubmissionID{tcom\_159s1}

\begin{document}

\title{UltraDiff: Differentiable Ray Tracing in Ultrasound for Shape Optimization}

\settopmatter{authorsperrow=4}
\author{Felix Duelmer}
\correspondingauthor
\affiliation{%
  \institution{Technical University of Munich \& Munich Center for Machine Learning}
  \city{Munich}
  \country{Germany}}
\email{felix.duelmer@tum.de}

\author{Magdalena Wysocki}
\affiliation{%
  \institution{Technical University of Munich \& Munich Center for Machine Learning}
  \city{Munich}
  \country{Germany}}

\author{Nassir Navab}
\affiliation{%
  \institution{Technical University of Munich \& Munich Center for Machine Learning}
  \city{Munich}
  \country{Germany}}

\author{Mohammad Farid Azampour}
\affiliation{%
  \institution{Technical University of Munich \& Munich Center for Machine Learning}
  \city{Munich}
  \country{Germany}}

\renewcommand{\shortauthors}{Duelmer et al.}

\begin{abstract}

Physically-based differentiable rendering enables gradient-based optimization of scene parameters by matching rendered images to measurements, but has so far mainly focused on light transport. 
We extend this paradigm to medical ultrasound, where image formation resembles transient rendering: echoes are binned by time-of-flight rather than projected onto an image plane. 
We present UltraDiff, a modular framework for differentiable ultrasound ray tracing. 
UltraDiff formulates ultrasound image formation as a path-space integral, gated by travel time between the transducer and tissue interfaces, and derives a Monte Carlo estimator of both the forward model and its gradients with respect to scene parameters. 
We demonstrate this on an inverse geometry estimation: starting from a sphere, an SDF is optimized until simulated echoes match measured ones, recovering vertebral surfaces from simulated B-mode sweeps and from a real robotic acquisition of a spine phantom. 
Unlike state-of-the-art ultrasound shape reconstruction methods, which rely on pre-segmented images, our approach operates unsupervised on B-mode images through analysis-by-synthesis, while achieving competitive geometric accuracy. 
Implemented on top of Mitsuba~3, UltraDiff brings differentiable path tracing to a new sensing modality and provides a foundation for inverse problems in acoustic imaging.
\end{abstract}


  
\begin{teaserfigure}
    \centering
    \includegraphics[width=0.95\textwidth]{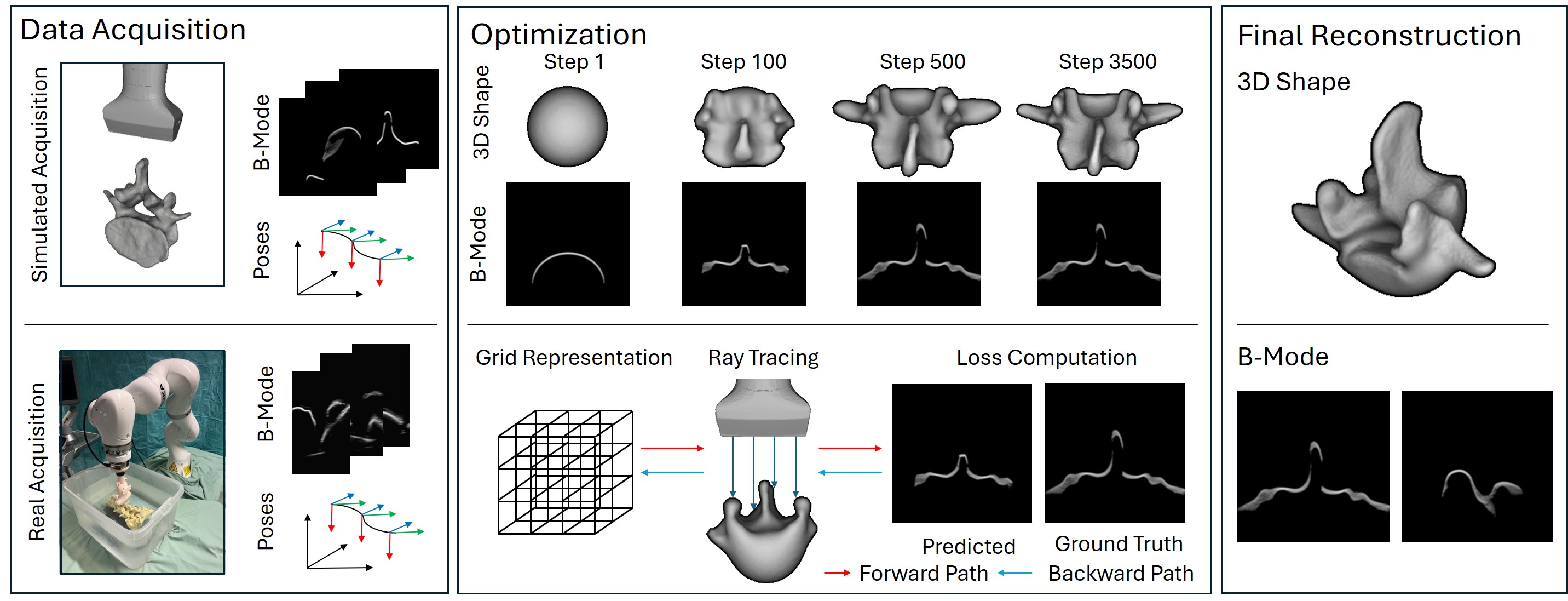}
    \caption{Shape optimization using UltraDiff. First, we acquire reference B-mode images and matching poses through simulation or real acquisition with robotic ultrasound. Next, we iteratively produce B-mode images based on the current scene state and our differentiable ray tracing framework, compute the loss between these simulated images and the target images, and backpropagate the loss through the ray tracing to optimize the underlying shape. }
    \label{fig:model_arch}
   \Description{Pipeline diagram: reference B-mode images and poses from simulation or robotic acquisition are compared with images rendered from the current SDF. The loss is backpropagated through the ray tracer to update the shape.}
\end{teaserfigure}

\maketitle

\section{Introduction} 
Physically-based differentiable rendering has become a powerful tool for inverse problems: by differentiating a Monte Carlo estimate of the rendering equation with respect to scene parameters, geometry and materials can be recovered from images via gradient descent \cite{li2018differentiable, zhang2020path, vicini2022differentiable}.
While most of this work targets light transport, the underlying machinery, specifically path-space integrals, Monte Carlo estimation, and boundary-aware gradients, is not specific to light. 
Time-resolved variants have been explored in transient rendering, where contributions are binned by time-of-flight rather than projected onto an image plane \cite{yi2021differentiable, wu2021differentiable}.

Medical ultrasound is, in this sense, a transient imaging modality: a transducer emits acoustic pulses, and echoes returning from tissue interfaces are placed in the image according to their travel time.
Recovering 3D anatomy from such images, e.g., bone surfaces from tracked B-mode sweeps, is a long-standing inverse problem in ultrasound-guided intervention. 
Yet state-of-the-art reconstruction methods \cite{chen2024rocosdf, wu2025ultraboneudf} do not invert the imaging process: they fit surfaces to point clouds extracted from pre-segmented images, tying accuracy to a separately trained segmentation model rather than to the measured intensities.
Differentiable forward models of ultrasound image formation offer an alternative: a scene representation is optimized until simulated images match the measured ones, inverting the imaging process directly.
Realizing this requires a forward model that is both differentiable and efficient.
Full waveform inversion solves the wave equation directly \cite{guasch2020full}, but differentiating full-wave solvers is computationally expensive and limits scalability.
Ray-based approximations \cite{mattausch2016monte, duelmer2025ultraray} are efficient and have recently been made differentiable \cite{wysocki2024ultra, duelmer2026ultrag}, but only for single-direction ray integration: each ray marches into the scene and accumulates along a straight line.
This limits the set of propagation paths that can contribute to the received echo and precludes view-dependent effects such as specular reflections and multi-bounce reverberation, which path tracing handles naturally.

In this paper, we therefore introduce UltraDiff, a differentiable ray tracing framework for ultrasound based on Monte Carlo path sampling.
UltraDiff formulates B-mode image formation as a time-of-flight gated path integral across multiple interactions with tissue boundaries and provides gradients of the rendered image with respect to scene parameters, including geometry represented as a signed distance field.
Implemented in Mitsuba~3, it enables segmentation-free vertebra reconstruction by analysis-by-synthesis from simulated and real robotic B-mode sweeps, competitive with segmentation-based methods.
Code and data are available at \url{https://github.com/Felixduelmer/ultradiff}.

\section{Method}

\subsection{Time-of-Flight Path Integral}
A B-mode ultrasound image is formed by emitting acoustic pulses from a transducer and recording returning echoes. 
Unlike a camera, which projects the scene onto an image plane, the position of an echo in the image is determined by its round-trip travel time.
Image formation is thus closely related to transient rendering \cite{yi2021differentiable, wu2021differentiable}, with the transducer acting as both emitter and sensor. 
We express the time-resolved pressure at transducer element $e$ as a path integral:
\begin{equation}\label{eq:td_pressure}
    P(e,t) = \int_{\Omega_e} f(\bar{x},t)\, d\mu(\bar{x})
    = \int_{\Omega_e} \mathfrak{T}(\bar{x})\,
    S_e(\bar{x},t)\, d\mu(\bar{x}),
\end{equation}
where $\Omega_e$ is the set of paths $\bar{x}=(x_0,\ldots,x_k)$ that start and end on the aperture at element $e$, with $k-1$ interaction points on tissue interfaces, and $d\mu$ is the path-space measure.
The integrand factors into a spatial throughput $\mathfrak{T}$ and a temporal response $S_e$.
The throughput collects the interaction terms $\rho(x_i)$, geometry factors $G$, and visibility $V$:
\begin{equation}\label{eq:throughput}
    \mathfrak{T}(\bar{x}) =
    \left[\prod_{i=1}^{k-1} \rho(x_i)\right]
    \left[\prod_{i=0}^{k-1} G(x_i,x_{i+1})\, V(x_i,x_{i+1})\right],
\end{equation}
so that richer acoustic interaction models can be substituted without changing the estimator. 
The temporal response gates each path by its time of flight $\textit{tof}(\bar{x}) = \frac{1}{\textit{SoS}}\sum_{i=0}^{k-1} \|x_{i+1}-x_i\|$, with $\textit{SoS}$ the speed of sound.
We instantiate $S_e$ as a Gaussian in $\textit{tof}(\bar{x})-t$ whose width is set by the transmitted pulse length: it models the temporal envelope of the transducer's pulse-echo response, governing axial resolution, and smoothly distributes each path's energy over neighboring time samples.
Drawing $N$ paths from a proposal density $p$ yields the standard unbiased estimator:
\begin{equation}\label{eq:mc_pressure}
P(e,t) \approx \frac{1}{N}\sum_{j=1}^{N}
\frac{\mathfrak{T}(\bar{x}_j)\, S_e(\bar{x}_j,t)}{p(\bar{x}_j)}.
\end{equation}

\subsection{Differentiable Formulation}
To optimize scene parameters $\theta$ (e.g., geometric or tissue parameters), we differentiate Eq.~\ref{eq:td_pressure}. 
Beyond the explicit dependence of the integrand on $\theta$, perturbing the geometry moves the interaction vertices themselves. 
Applying the transport theorem to the evolving domain yields:
\begin{equation}\label{eq:diff}
\frac{d}{d\theta} P(e,t) = \int_{\Omega_e} 
\left[
\frac{\partial f(\bar{x},t)}{\partial \theta}
-
f(\bar{x},t)\sum_{i=1}^{k-1}
\mathbf{v}(x_i)\cdot \mathbf{n}(x_i)
\right] d\mu(\bar{x}),
\end{equation}
where $\mathbf{n}(x_i)$ is the surface normal and $\mathbf{v}(x_i)$ the boundary velocity induced by $\theta$, analogous to the normal-velocity term in differentiable SDF rendering \cite{vicini2022differentiable}. 
We estimate the gradient with the same samples as Eq.~\ref{eq:mc_pressure}. 
Note that we do not explicitly sample boundary terms arising from visibility discontinuities \cite{li2018differentiable,
zhang2020path}. In our setting, the Gaussian temporal kernel and the diffuse extent of the transducer response smooth the dominant discontinuities, which we find sufficient for stable optimization.

\subsection{Shape Representation}

We represent geometry as a signed distance function (SDF) on a voxel grid with cubic B-spline interpolation, which provides smooth gradients for optimization \cite{vicini2022differentiable, wang2024simple}. 
Intersections are found by sphere tracing \cite{hart1996sphere}. Velocities and normals in Eq.~\ref{eq:diff} follow directly from the SDF: $\mathbf{v}(x) = -\partial_\theta SDF(x)$ and $\mathbf{n}(x) = \nabla SDF(x)/\lVert \nabla SDF(x) \rVert$.

\subsection{Implementation}
We implement UltraDiff in Mitsuba~3 using its modular emitter and integrator interfaces, running on the CUDA backend via Dr.Jit \cite{jakob2022dr}.
For the shape reconstruction experiments, we restrict paths to a single tissue interaction ($x_0 \to x_1 \to x_2$, with $x_0 \equiv x_2$), which captures specular echoes from large-scale tissue boundaries and keeps the inverse problem well-conditioned. 
The path-space formulation itself is not limited to one bounce, and we demonstrate two-bounce transport in Sec.~\ref{sec:gradients}.
We estimate Eq.~\ref{eq:mc_pressure} by drawing $N$ rays sampled uniformly on each element~$e$.
Following common practice in transient rendering~\cite{yi2021differentiable}, the same rays contribute to all time samples: we splat each ray's contribution into neighboring bins according to its travel distance via~$S_e$.
The SDF is optimized with Adam under an $\ell_1$ image loss, using a coarse-to-fine schedule ($64^3$ to $512^3$), Laplacian regularization, and periodic re-distancing by fast sweeping \cite{vicini2022differentiable}. 
On an NVIDIA RTX~4070~Ti, optimization traces $\sim$10M rays in parallel at $\approx$10 epochs/s and completes in roughly 10 minutes.

\section{Experiments and Results}

\subsection{Gradient Validation and Multi-Bounce Transport}
\label{sec:gradients}
\begin{figure}[h]
    \centering
    \includegraphics[width=0.9\linewidth]{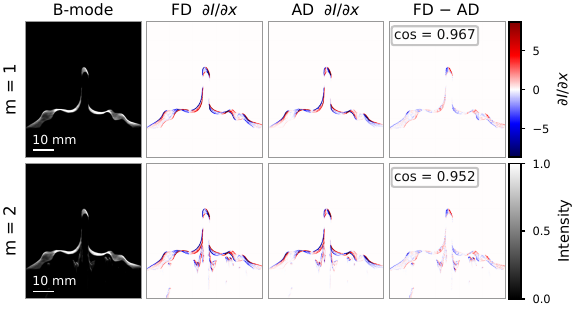}
    \caption{Gradient validation with respect to a lateral SDF translation for single-bounce (top, $m{=}1$) and two-bounce (bottom, $m{=}2$) transport. 
Left to right: rendered B-mode, finite-difference gradient (FD), our analytic gradient (AD), and their difference. 
The two-bounce row shows a ghost echo at greater apparent depth caused by multi-path transport.}
    \Description{A two-by-four grid of ultrasound images of a vertebra. Each of the two rows shows, from left to right, a grayscale B-mode image, two gradient images in a red--blue diverging colormap with closely matching band structures along the bone boundary, and a difference image that is nearly blank. 
    The bottom (two-bounce) row shows additional faint echoes below the bone surface.}
    \label{fig:gradients}
\end{figure}

We validate our gradients on a vertebra scene. We perturb the shape through a single parameter, a lateral translation of the signed distance field (SDF), and compare the analytic gradient of the rendered B-mode image against a finite-difference reference (Fig.~\ref{fig:gradients}). 
The analytic (AD) and finite-difference (FD) gradients agree closely, with cosine similarities of $0.97$ and $0.95$ for single- and two-bounce transport, respectively. 
The gradient concentrates in a thin, signed band along the reflecting tissue boundary, where the round-trip echo originates, rather than along projected silhouettes as in optical differentiable rendering, a direct consequence of differentiating through the time-of-flight gating of the echo signal.
The residual (Fig.~\ref{fig:gradients}, right column) is mostly confined to sharp boundary curvature.
We further show that the path-space formulation extends beyond single-bounce transport.
Tracing paths with two interactions reproduces multi-path artifacts, commonly observed in real ultrasound: an echo reflects off one interface onto a second before returning to the transducer. Because each echo is placed by its round-trip time, this signal appears as a ghost echo at greater apparent depth than either reflecting surface. Such paths exist only where a specular bounce can reach a second surface, i.e., within concave anatomy, and are structurally unreachable for single-direction ray-integration models~\cite{wysocki2024ultra, duelmer2026ultrag}.

\subsection{Shape Reconstruction}\label{sec:shape_recon}

\paragraph{Datasets}
We evaluate shape optimization on two datasets: a synthetic dataset of five vertebrae from VerSe2020 \cite{sekuboyina2021verse} (CC BY-SA 4.0) rendered with our forward model, and a real robotic ultrasound acquisition of a spine phantom in a water bath.
For the synthetic data, we simulate three sweeps per vertebra (two longitudinal, one transverse), each at three
out-of-plane tilt angles of $-10^\circ$, $0^\circ$, and $10^\circ$, yielding 450 posed B-mode images per vertebra at 6~cm imaging depth. 
The real acquisition consists of a single sweep at three tilt angles of $-20^\circ$, $0^\circ$, and $20^\circ$ at 9~cm depth, yielding $\sim$600 tracked images. 
In both settings we use the same transducer model with an isotropic point spread function corresponding to a 5~MHz probe. 
For the real data, we discard the lower 30\% of each image, which is dominated by low-SNR clutter that our specular-reflection forward model does not represent.

\paragraph{Baselines}
We compare against three recent ultrasound shape reconstruction methods: RoCoSDF \cite{chen2024rocosdf}, which fits neural SDFs to orthogonal sweeps and fuses them in a second stage. 
UltraBoneUDF \cite{wu2025ultraboneudf}, which reconstructs an unsigned distance field of the bone surface, and UltrON \cite{wysocki2025ultron}, an occupancy-based approach. 
All three require point clouds extracted from segmented B-mode images as input. 
We provide these by treating each non-zero pixel of our processed target images as a 3D point.
Since the processed targets contain only bone-surface echoes, this acts as an idealized segmentation that gives the baselines the same observations as UltraDiff.

\paragraph{Metrics}
We report Chamfer distance, mean absolute surface deviation (MAD), earth mover's distance (EMD), and the 95th-percentile Hausdorff distance (HD95), all in mm, computed only on surface regions observable from the transducer poses so that occluded geometry does not dominate the error.

\begin{table}[h]
\centering
\caption{Geometric error on the synthetic dataset (mm, mean$\pm$std over five vertebrae). 
Best values in bold.}
\label{tab:shape_evaluation_mm}
\setlength{\tabcolsep}{3.5pt}
\small
\begin{tabular*}{\linewidth}{@{\extracolsep{\fill}}l c c c c @{}}
\toprule
\textbf{Method}
& \textbf{Chamfer} $\downarrow$
& \textbf{HD95} $\downarrow$
& \textbf{MAD} $\downarrow$
& \textbf{EMD} $\downarrow$ \\
\midrule
RoCoSDF
  & 3.36{\scriptsize$\pm$0.14}
  & \textbf{2.31}{\scriptsize$\pm$0.05}
  & 1.66{\scriptsize$\pm$0.08}
  & 3.22{\scriptsize$\pm$0.17} \\
UltraBoneUDF
  & 3.04{\scriptsize$\pm$0.25}
  & 2.78{\scriptsize$\pm$0.66}
  & 1.50{\scriptsize$\pm$0.16}
  & 3.62{\scriptsize$\pm$0.38} \\
UltrON
  & 3.33{\scriptsize$\pm$0.40}
  & 4.55{\scriptsize$\pm$2.78}
  & 1.36{\scriptsize$\pm$0.07}
  & 4.41{\scriptsize$\pm$0.70} \\
UltraDiff
  & \textbf{0.79}{\scriptsize$\pm$0.09}
  & 3.20{\scriptsize$\pm$0.62}
  & \textbf{0.20}{\scriptsize$\pm$0.05}
  & \textbf{2.87}{\scriptsize$\pm$0.28} \\
\bottomrule
\end{tabular*}
\end{table}

\begin{figure}[h]
    \centering
    \includegraphics[width=0.9\linewidth]{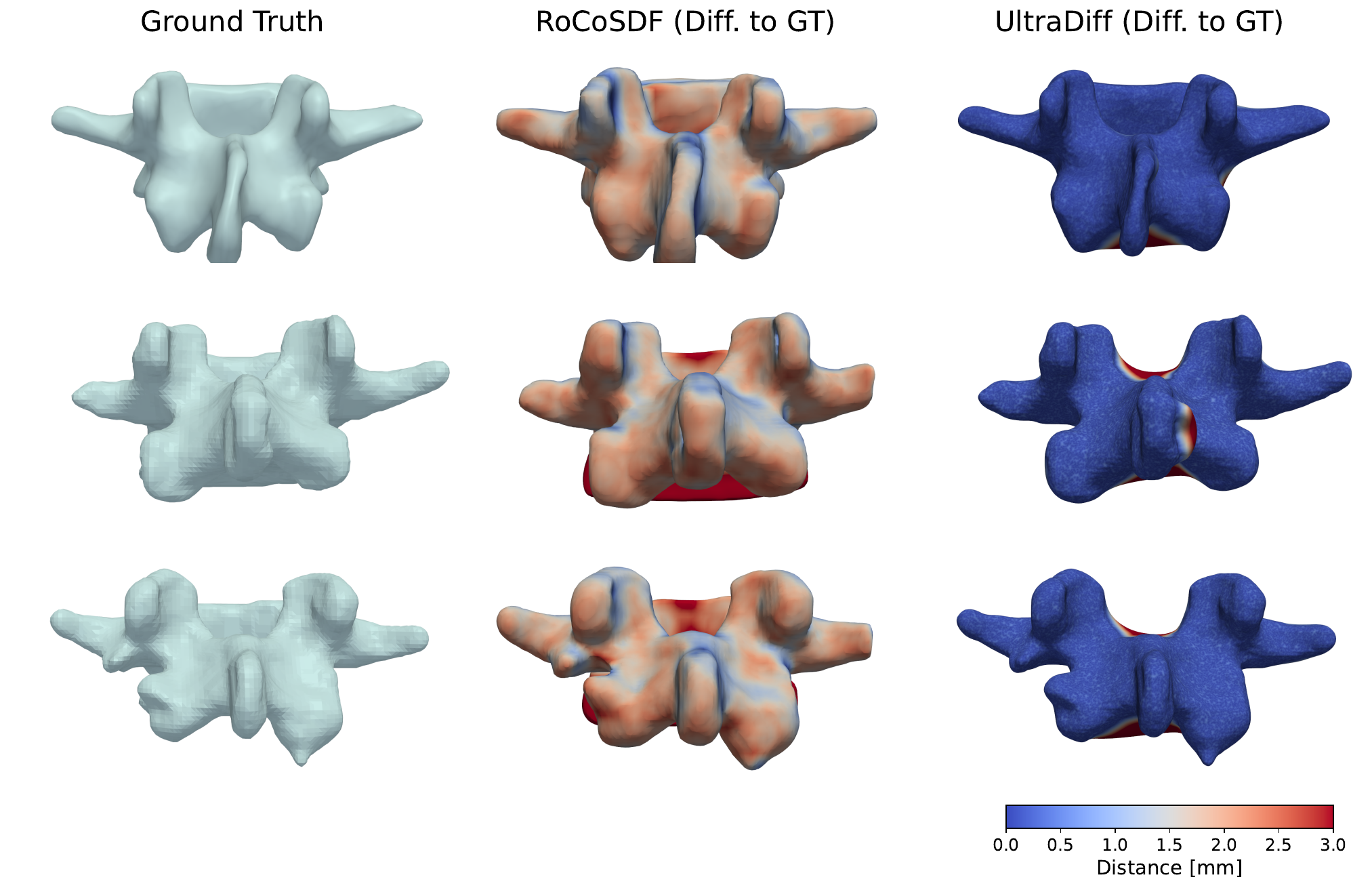}
    \caption{Reconstructions on the synthetic dataset. 
        Rows show sample vertebrae.
        Columns show the ground-truth surface, RoCoSDF~\cite{chen2024rocosdf}, and UltraDiff. 
        Color encodes per-point distance to the ground truth in mm.}
 \Description{A grid of rendered vertebra surfaces, three per row, shown
      from a consistent viewpoint. The left column shows the ground-truth
      surfaces in a neutral shade. The middle and right columns show
      reconstructions of the same vertebrae shaded by a color scale running
      from blue at zero error to red at the maximum. The middle (RoCoSDF)
      surfaces are broadly green-to-yellow across the vertebral body and
      processes, with the spinous and transverse processes noticeably
      blunted or truncated relative to the ground truth. The right
      (UltraDiff) surfaces are predominantly blue, closely following the
      ground-truth outline including the process tips, with isolated
      yellow-to-red patches confined to the intervertebral disc region.}
    \label{fig:comparison_shapes}
\end{figure}

\paragraph{Results}

\begin{figure}[h]
    \centering
    \includegraphics[width=0.81\linewidth]{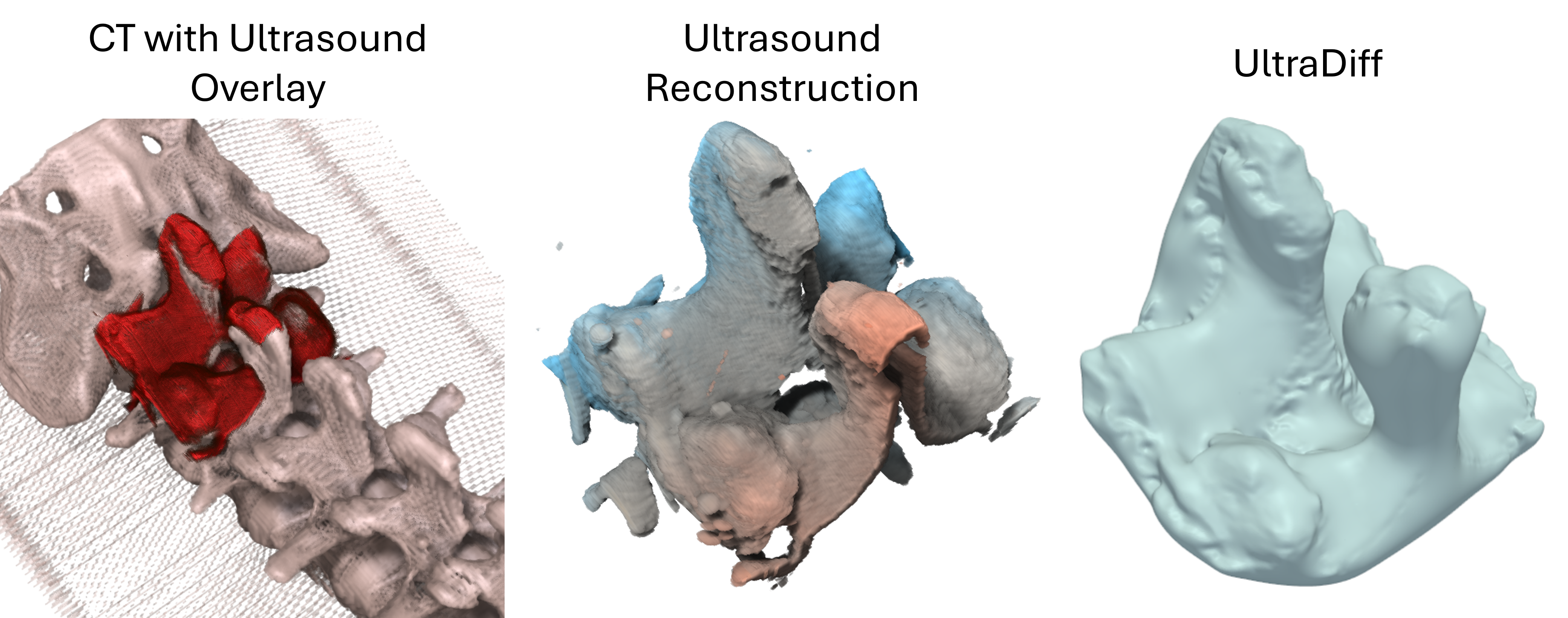}
    \caption{Real spine-phantom reconstruction. Left: phantom CT with imaged region. Middle: full reconstruction (manual color map). Right: reconstructed vertebrae.}
    \Description{Three panels. The left panel is a CT rendering of the spine
  phantom with the imaged region marked. The middle panel shows the full
  three-dimensional reconstruction rendered with a manually chosen color
  map. The right panel shows the vertebral surfaces recovered by
  UltraDiff, closely following the shape of the corresponding vertebrae
  in the CT.}
    \label{fig:comparison_shapes_real}
\end{figure}

On the synthetic dataset, UltraDiff reduces Chamfer distance by $3.8\times$ and MAD by $6.8\times$ relative to the best baseline (Table~\ref{tab:shape_evaluation_mm}), and the reconstructions closely match the ground-truth surfaces (Fig.~\ref{fig:comparison_shapes}). 
The main failure case is the intervertebral disc region, which is observed in only a subset of views: there, the Laplacian prior can over-regularize and suppress the disc, producing localized outliers that raise HD95.
On the real phantom acquisition (Fig.~\ref{fig:comparison_shapes_real}), the same forward model and optimization settings transfer without modification. 
UltraDiff recovers the overall anatomy and principal curvature reliably, with remaining deviations localized to regions of weak or ambiguous signal.
Our sub-millimeter MAD lies within the lumbar tolerances for image-guided pedicle screw placement (up to 3.8\,mm \cite{rampersaud2001accuracy}).

\section{Discussion and Conclusion}

We introduced UltraDiff, a differentiable ray tracing framework that extends physically-based rendering to ultrasound. 
By formulating B-mode image formation as a time-of-flight gated path integral, we obtain a Monte Carlo estimator of both the forward model and its gradients, enabling end-to-end optimization through the simulator. 
The framework is intentionally modular: transducer models, interaction terms, and priors can be exchanged with minimal changes to the pipeline, and the path-space formulation supports multi-bounce transport, as demonstrated by the multi-path artifacts in Sec.~\ref{sec:gradients}.
Our shape reconstruction results indicate that the forward model yields well-behaved gradients for geometric refinement: operating directly on B-mode images without segmentation, UltraDiff matches or exceeds the accuracy of point-cloud-based state-of-the-art methods and transfers to real acquisition data without modification. 
Several limitations remain. 
Our gradient estimator does not explicitly sample boundary terms from visibility discontinuities \cite{li2018differentiable, zhang2020path}. 
Structures only observed in a few views, such as the intervertebral disc, can be suppressed by the smoothness prior (Sec.~\ref{sec:shape_recon}), suggesting schedules that begin with a broader point spread function and progressively sharpen. 
Finally, our synthetic targets are rendered with our own forward model, which may bias the comparison in our favor. 
The direct transfer to real data mitigates, but does not eliminate, this concern.
UltraDiff brings differentiable path tracing to a new sensing modality. 
We see it as a foundation for richer acoustic interaction models, namely, scattering and attenuation, as well as for broader inverse problems, such as transducer and acquisition trajectory optimization. 

\bibliographystyle{ACM-Reference-Format}
\bibliography{sample-base}


\begin{thebibliography}{18}


\ifx \showCODEN    \undefined \def \showCODEN     #1{\unskip}     \fi
\ifx \showISBNx    \undefined \def \showISBNx     #1{\unskip}     \fi
\ifx \showISBNxiii \undefined \def \showISBNxiii  #1{\unskip}     \fi
\ifx \showISSN     \undefined \def \showISSN      #1{\unskip}     \fi
\ifx \showLCCN     \undefined \def \showLCCN      #1{\unskip}     \fi
\ifx \shownote     \undefined \def \shownote      #1{#1}          \fi
\ifx \showarticletitle \undefined \def \showarticletitle #1{#1}   \fi
\ifx \showURL      \undefined \def \showURL       {\relax}        \fi
\providecommand\bibfield[2]{#2}
\providecommand\bibinfo[2]{#2}
\providecommand\natexlab[1]{#1}
\providecommand\showeprint[2][]{arXiv:#2}

\bibitem[Chen et~al\mbox{.}(2024)]%
        {chen2024rocosdf}
\bibfield{author}{\bibinfo{person}{Hongbo Chen}, \bibinfo{person}{Yuchong Gao},
  \bibinfo{person}{Shuhang Zhang}, \bibinfo{person}{Jiangjie Wu},
  \bibinfo{person}{Yuexin Ma}, {and} \bibinfo{person}{Rui Zheng}.}
  \bibinfo{year}{2024}\natexlab{}.
\newblock \showarticletitle{RoCoSDF: Row-Column Scanned Neural Signed Distance
  Fields for Freehand 3D Ultrasound Imaging Shape Reconstruction}. In
  \bibinfo{booktitle}{\emph{International Conference on Medical Image Computing
  and Computer-Assisted Intervention}}. Springer, \bibinfo{pages}{721--731}.
\newblock


\bibitem[Duelmer et~al\mbox{.}(2025)]%
        {duelmer2025ultraray}
\bibfield{author}{\bibinfo{person}{Felix Duelmer},
  \bibinfo{person}{Mohammad~Farid Azampour}, \bibinfo{person}{Magdalena
  Wysocki}, {and} \bibinfo{person}{Nassir Navab}.}
  \bibinfo{year}{2025}\natexlab{}.
\newblock \showarticletitle{Ultraray: Introducing full-path ray tracing in
  physics-based ultrasound simulation}. In
  \bibinfo{booktitle}{\emph{International Conference on Medical Image Computing
  and Computer-Assisted Intervention}}. Springer, \bibinfo{pages}{653--662}.
\newblock


\bibitem[Duelmer et~al\mbox{.}(2026)]%
        {duelmer2026ultrag}
\bibfield{author}{\bibinfo{person}{Felix Duelmer}, \bibinfo{person}{Jakob
  Klaushofer}, \bibinfo{person}{Magdalena Wysocki}, \bibinfo{person}{Nassir
  Navab}, {and} \bibinfo{person}{Mohammad~Farid Azampour}.}
  \bibinfo{year}{2026}\natexlab{}.
\newblock \showarticletitle{UltraG-Ray: Physics-Based Gaussian Ray Casting for
  Novel Ultrasound View Synthesis}. In \bibinfo{booktitle}{\emph{Medical
  Imaging with Deep Learning}}.
\newblock


\bibitem[Guasch et~al\mbox{.}(2020)]%
        {guasch2020full}
\bibfield{author}{\bibinfo{person}{Llu{\'\i}s Guasch}, \bibinfo{person}{Oscar
  Calder{\'o}n~Agudo}, \bibinfo{person}{Meng-Xing Tang},
  \bibinfo{person}{Parashkev Nachev}, {and} \bibinfo{person}{Michael Warner}.}
  \bibinfo{year}{2020}\natexlab{}.
\newblock \showarticletitle{Full-waveform inversion imaging of the human
  brain}.
\newblock \bibinfo{journal}{\emph{NPJ digital medicine}} \bibinfo{volume}{3},
  \bibinfo{number}{1} (\bibinfo{year}{2020}), \bibinfo{pages}{28}.
\newblock


\bibitem[Hart(1996)]%
        {hart1996sphere}
\bibfield{author}{\bibinfo{person}{John~C Hart}.}
  \bibinfo{year}{1996}\natexlab{}.
\newblock \showarticletitle{Sphere tracing: A geometric method for the
  antialiased ray tracing of implicit surfaces}.
\newblock \bibinfo{journal}{\emph{The Visual Computer}} \bibinfo{volume}{12},
  \bibinfo{number}{10} (\bibinfo{year}{1996}), \bibinfo{pages}{527--545}.
\newblock


\bibitem[Jakob et~al\mbox{.}(2022)]%
        {jakob2022dr}
\bibfield{author}{\bibinfo{person}{Wenzel Jakob},
  \bibinfo{person}{S{\'e}bastien Speierer}, \bibinfo{person}{Nicolas Roussel},
  {and} \bibinfo{person}{Delio Vicini}.} \bibinfo{year}{2022}\natexlab{}.
\newblock \showarticletitle{Dr. jit: A just-in-time compiler for differentiable
  rendering}.
\newblock \bibinfo{journal}{\emph{ACM Transactions on Graphics (TOG)}}
  \bibinfo{volume}{41}, \bibinfo{number}{4} (\bibinfo{year}{2022}),
  \bibinfo{pages}{1--19}.
\newblock


\bibitem[Li et~al\mbox{.}(2018)]%
        {li2018differentiable}
\bibfield{author}{\bibinfo{person}{Tzu-Mao Li}, \bibinfo{person}{Miika
  Aittala}, \bibinfo{person}{Fr{\'e}do Durand}, {and} \bibinfo{person}{Jaakko
  Lehtinen}.} \bibinfo{year}{2018}\natexlab{}.
\newblock \showarticletitle{Differentiable monte carlo ray tracing through edge
  sampling}.
\newblock \bibinfo{journal}{\emph{ACM Transactions on Graphics (TOG)}}
  \bibinfo{volume}{37}, \bibinfo{number}{6} (\bibinfo{year}{2018}),
  \bibinfo{pages}{1--11}.
\newblock


\bibitem[Mattausch and Goksel(2016)]%
        {mattausch2016monte}
\bibfield{author}{\bibinfo{person}{Oliver Mattausch} {and}
  \bibinfo{person}{Orcun Goksel}.} \bibinfo{year}{2016}\natexlab{}.
\newblock \showarticletitle{Monte-carlo ray-tracing for realistic interactive
  ultrasound simulation}. In \bibinfo{booktitle}{\emph{Proceedings of the
  Eurographics Workshop on Visual Computing for Biology and Medicine}}.
  \bibinfo{pages}{173--181}.
\newblock


\bibitem[Rampersaud et~al\mbox{.}(2001)]%
        {rampersaud2001accuracy}
\bibfield{author}{\bibinfo{person}{Y~Raja Rampersaud}, \bibinfo{person}{David~A
  Simon}, {and} \bibinfo{person}{Kevin~T Foley}.}
  \bibinfo{year}{2001}\natexlab{}.
\newblock \showarticletitle{Accuracy requirements for image-guided spinal
  pedicle screw placement}.
\newblock \bibinfo{journal}{\emph{Spine}} \bibinfo{volume}{26},
  \bibinfo{number}{4} (\bibinfo{year}{2001}), \bibinfo{pages}{352--359}.
\newblock


\bibitem[Sekuboyina et~al\mbox{.}(2021)]%
        {sekuboyina2021verse}
\bibfield{author}{\bibinfo{person}{Anjany Sekuboyina}, \bibinfo{person}{Malek~E
  Husseini}, \bibinfo{person}{Amirhossein Bayat}, \bibinfo{person}{Maximilian
  L{\"o}ffler}, \bibinfo{person}{Hans Liebl}, \bibinfo{person}{Hongwei Li},
  \bibinfo{person}{Giles Tetteh}, \bibinfo{person}{Jan Kuka{\v{c}}ka},
  \bibinfo{person}{Christian Payer}, \bibinfo{person}{Darko {\v{S}}tern},
  {et~al\mbox{.}}} \bibinfo{year}{2021}\natexlab{}.
\newblock \showarticletitle{VerSe: a vertebrae labelling and segmentation
  benchmark for multi-detector CT images}.
\newblock \bibinfo{journal}{\emph{Medical image analysis}}
  \bibinfo{volume}{73} (\bibinfo{year}{2021}), \bibinfo{pages}{102166}.
\newblock


\bibitem[Vicini et~al\mbox{.}(2022)]%
        {vicini2022differentiable}
\bibfield{author}{\bibinfo{person}{Delio Vicini},
  \bibinfo{person}{S{\'e}bastien Speierer}, {and} \bibinfo{person}{Wenzel
  Jakob}.} \bibinfo{year}{2022}\natexlab{}.
\newblock \showarticletitle{Differentiable signed distance function rendering}.
\newblock \bibinfo{journal}{\emph{ACM Transactions on Graphics (TOG)}}
  \bibinfo{volume}{41}, \bibinfo{number}{4} (\bibinfo{year}{2022}),
  \bibinfo{pages}{1--18}.
\newblock


\bibitem[Wang et~al\mbox{.}(2024)]%
        {wang2024simple}
\bibfield{author}{\bibinfo{person}{Zichen Wang}, \bibinfo{person}{Xi Deng},
  \bibinfo{person}{Ziyi Zhang}, \bibinfo{person}{Wenzel Jakob}, {and}
  \bibinfo{person}{Steve Marschner}.} \bibinfo{year}{2024}\natexlab{}.
\newblock \showarticletitle{A Simple Approach to Differentiable Rendering of
  SDFs}. In \bibinfo{booktitle}{\emph{SIGGRAPH Asia 2024 Conference Papers}}.
  \bibinfo{pages}{1--11}.
\newblock


\bibitem[Wu et~al\mbox{.}(2021)]%
        {wu2021differentiable}
\bibfield{author}{\bibinfo{person}{Lifan Wu}, \bibinfo{person}{Guangyan Cai},
  \bibinfo{person}{Ravi Ramamoorthi}, {and} \bibinfo{person}{Shuang Zhao}.}
  \bibinfo{year}{2021}\natexlab{}.
\newblock \showarticletitle{Differentiable time-gated rendering}.
\newblock \bibinfo{journal}{\emph{ACM Transactions on Graphics (TOG)}}
  \bibinfo{volume}{40}, \bibinfo{number}{6} (\bibinfo{year}{2021}),
  \bibinfo{pages}{1--16}.
\newblock


\bibitem[Wu et~al\mbox{.}(2025)]%
        {wu2025ultraboneudf}
\bibfield{author}{\bibinfo{person}{Luohong Wu}, \bibinfo{person}{Matthias
  Seibold}, \bibinfo{person}{Nicola~A Cavalcanti}, \bibinfo{person}{Giuseppe
  Loggia}, \bibinfo{person}{Lisa Reissner}, \bibinfo{person}{Bastian Sigrist},
  \bibinfo{person}{Jonas Hein}, \bibinfo{person}{Lilian Calvet},
  \bibinfo{person}{Arnd Vieh{\"o}fer}, {and} \bibinfo{person}{Philipp
  F{\"u}rnstahl}.} \bibinfo{year}{2025}\natexlab{}.
\newblock \showarticletitle{UltraBoneUDF: Self-supervised bone surface
  reconstruction from ultrasound based on neural unsigned distance functions}.
\newblock \bibinfo{journal}{\emph{Computerized Medical Imaging and Graphics}}
  (\bibinfo{year}{2025}), \bibinfo{pages}{102690}.
\newblock


\bibitem[Wysocki et~al\mbox{.}(2024)]%
        {wysocki2024ultra}
\bibfield{author}{\bibinfo{person}{Magdalena Wysocki},
  \bibinfo{person}{Mohammad~Farid Azampour}, \bibinfo{person}{Christine
  Eilers}, \bibinfo{person}{Benjamin Busam}, \bibinfo{person}{Mehrdad Salehi},
  {and} \bibinfo{person}{Nassir Navab}.} \bibinfo{year}{2024}\natexlab{}.
\newblock \showarticletitle{Ultra-nerf: Neural radiance fields for ultrasound
  imaging}. In \bibinfo{booktitle}{\emph{Medical Imaging with Deep Learning}}.
  PMLR, \bibinfo{pages}{382--401}.
\newblock


\bibitem[Wysocki et~al\mbox{.}(2025)]%
        {wysocki2025ultron}
\bibfield{author}{\bibinfo{person}{Magdalena Wysocki}, \bibinfo{person}{Felix
  Duelmer}, \bibinfo{person}{Ananya Bal}, \bibinfo{person}{Nassir Navab}, {and}
  \bibinfo{person}{Mohammad~Farid Azampour}.} \bibinfo{year}{2025}\natexlab{}.
\newblock \showarticletitle{UltrON: Ultrasound Occupancy Networks}. In
  \bibinfo{booktitle}{\emph{International Conference on Medical Image Computing
  and Computer-Assisted Intervention}}. Springer, \bibinfo{pages}{606--615}.
\newblock


\bibitem[Yi et~al\mbox{.}(2021)]%
        {yi2021differentiable}
\bibfield{author}{\bibinfo{person}{Shinyoung Yi}, \bibinfo{person}{Donggun
  Kim}, \bibinfo{person}{Kiseok Choi}, \bibinfo{person}{Adrian Jarabo},
  \bibinfo{person}{Diego Gutierrez}, {and} \bibinfo{person}{Min~H Kim}.}
  \bibinfo{year}{2021}\natexlab{}.
\newblock \showarticletitle{Differentiable transient rendering}.
\newblock \bibinfo{journal}{\emph{ACM Transactions on Graphics (TOG)}}
  \bibinfo{volume}{40}, \bibinfo{number}{6} (\bibinfo{year}{2021}),
  \bibinfo{pages}{1--11}.
\newblock


\bibitem[Zhang et~al\mbox{.}(2020)]%
        {zhang2020path}
\bibfield{author}{\bibinfo{person}{Cheng Zhang}, \bibinfo{person}{Bailey
  Miller}, \bibinfo{person}{Kai Yan}, \bibinfo{person}{Ioannis Gkioulekas},
  {and} \bibinfo{person}{Shuang Zhao}.} \bibinfo{year}{2020}\natexlab{}.
\newblock \showarticletitle{Path-space differentiable rendering}.
\newblock \bibinfo{journal}{\emph{ACM Transactions on Graphics}}
  \bibinfo{volume}{39}, \bibinfo{number}{4} (\bibinfo{year}{2020}),
  \bibinfo{pages}{143--1}.
\newblock


\end{thebibliography}

\end{document}